\documentclass[a4paper,12pt]{article}

\usepackage{graphicx}

\newcommand{\beq}{\begin{eqnarray}}
\newcommand{\eeq}{\end{eqnarray}}

\newcommand{\ket}{\rangle}
\newcommand{\bra}{\langle}
\newcommand{\del}{\partial}

\newcommand{\dslash}{{\del \hspace{-5pt}/}}

\def\gsim{\displaystyle\mathop{>}_{\sim}}
\def\lsim{\displaystyle\mathop{<}_{\sim}}

\begin{document}

\begin{center}
{\large  \bf
Pentaquark states in a chiral potential}
 
 \vspace*{1cm}
 Atsushi Hosaka\\
 \vspace*{2mm}
 {\it Research Center for Nuclear Physics (RCNP), Osaka University\\
 Ibaraki 567-0047 Japan}
\end{center}
\vspace*{0.2cm}

\abstract{
We discuss a possible interpretation for 
the pentaquark baryon $\Theta^+$ which 
has an exotic quantum number of strangeness $S = +1$.  
The role of the  pion which modifies the structure of 
single-particle levels of quarks  is examined.  
When the  pion field is sufficiently strong, the ground state of five quarks 
for $\Theta^+$ 
acquires positive parity.  
If this is the case, an excited state may appear slightly 
above the ground state with negative parity.}

\vspace*{1cm}

Recently, in the measurement of $K^{\pm}$ in a photo-induced reaction from 
the neutron, $\gamma n ({\rm  in} ^{12}C) \to K^+ K^- n$, Nakano and 
collaborators (LEPS at SPring-8) has observed a sharp peak in the $K^+ n$ 
invariant mass analysis~\cite{LEPS}.  
It is located  at $M \sim 1540 \pm 10$ MeV with a very narrow 
width $\Gamma \lsim 25$ MeV.  
Subsequently,  DIANA collaboration at ITEP~\cite{DIANA}  
and CLAS collaboration at Jefferson Lab~\cite{CLAS} 
have also confirmed the existence of the resonance structure.   
The state corresponding to this sharp peak carries unit charge $Q = +1$ and 
positive strangeness $S = + 1$ as a 
composite system of $K^+ n$, and  has been identified with an exotic baryon 
which is now denoted as  $\Theta^+$.  

\begin{figure}
       \centerline{\includegraphics[width=5cm]
                                   {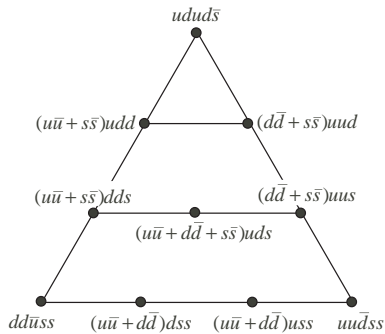}}
\centering
\begin{minipage}{12cm}
   \caption{\small 
   A wight diagram for antidecuplet $\bar {10}$ formed by 
   using two flavor conjugate diquarks $(\bar q)_{qq}$ 
   and one antiquark $\bar q$.  Here $q \bar q$ denotes
$u \bar u$, $d \bar d $ and  $s \bar s$.    }
   \label{antidec}
 \end{minipage}
\end{figure}

In a naive valence quark model, such a state can not be described as a 
conventional three quark state.  
The flavor quantum numbers of ordinary $qqq$ baryons are 
dictated by the SU(3) irreducible decomposition, 
$3 \times 3 \times 3 = 1 + 8 + 8 + 10$.  
As a consequence, the strangeness of $qqq$  baryons are limited to 
$S = -3$  ($\Omega$),  $-2$  ($\Xi$), 
$-1$  ($\Sigma$,  $\Lambda$) and  $0$ ($N$). 
Hence the $S=+1$ state must require at least five quarks 
($qqqq \bar q$).  
Such a state is a member of antidecuplet $\bar {10}$ or even 
higher dimensional representations such as 27-let.  
Using  flavor conjugate diquarks 
$(\bar q_i )_{qq} \equiv \epsilon_{ijk} q_j q_k$, 
$\Theta^+$ in the antidecuplet can be constructed as 
\beq
|\Theta^+\ket \sim 
(\bar s)_{qq} (\bar s)_{qq} \bar s \, , \; \; \; \; \; 
(\bar s)_{qq} \sim ud \, .
\eeq
The members of the antidecuplet are shown in Fig.~\ref{antidec}.  
From this construction, the isospin of $\Theta^+$ is $I = 0$.  
In $\bar {10}$, the seven baryons located at the cites 
other than the three corners of the triangle may be affected by  mixing 
with  octet members.  
However, the three states on the corners are genuine exotic pentaquark  
states.

Theoretically, using the chiral quark model, 
Diakonov et al~\cite{diakonov} have already 
studied  the properties of $\Theta^+$ as a member of antidecuplet 
with  predictions of masses and width 
which are surprisingly very close to what the experimentalists observed.  
Furthermore, the spin and parity $J^P$ were predicted to be 
$J^P = 1/2^+$.  
Once  the
theoretical method of the chiral quark model are accepted, 
these consequences can  be derived in a straightforward manner.  
In particular, many of the
mass formulae and relations among decay widths are dictated by the
symmetry (algebraic) relations~\cite{walliser}.  
However, if we attempt to understand these properties in the naive 
quark model, we will be immediately in trouble, since even a very 
fundamental quantum number such as parity does not agree with the 
prediction of the chiral quark model.  
Since the lowest quark state is the $s$ state ($l=0$), 
the lowest  $qqqq\bar q$  state must have negative parity.  
In another picture of a meson-baryon hybrid (molecule), 
a $K^{+}n$ (or $K^0 p$) bound state in an $s$ orbit also
carries negative parity.

In the latter approach, however, it has been known that the
$p$ state rather than the $s$ state becomes lowest for
$KN$ (=$K^{+}n$ and $K^{0}p$) and $\bar K N$ systems 
in the bound state approach of
the Skyrme model~\cite{callan}.
There, the kaon interaction with the hedgehog soliton plays a
peculiar but crucially important role to change the ordering of
the $s$ and $p$ states of the kaon. 
As a consequence, the state of $J^{P}=1/2^{+}$
appears as the lowest state both for the $S=\pm 1$ sector.
Furthermore, these states are split by the Wess-Zumino-Witten term
such that the $S=-1$ sector is lowered, while the 
$S=+1$ state is pushed up.  
Now comparing the Skyrme model and the chiral quark soliton model, they
must have a common feature which has the origin in the
non-perturbative chiral dynamics, since the starting point of
their methods is the formation of the hedgehog soliton configuration. 
The strong correlation generated  by the pion must
have the crucial role.  
In fact, in a recent report, Stancu and Riska studied   
quark energies of the $uu dd \bar s$ states  in  
a chiral quark model~\cite{stancu}.  
It was shown that flavor dependent interaction through   
Nambu-Goldstone boson (pion) exchanges affected strongly the structure of the 
quark levels and interchanged the ordering of the $s$ and $p$ 
orbits.  

In this paper, we  show that the role of the pion is understood  clearly 
by varying the strength of the 
interaction and by seeing the structure of the single particle levels of quarks.  
This is most conveniently demonstrated  in the chiral bag 
model~\cite{cbag}, where 
an interaction  of the $\vec \sigma \cdot \vec \tau$ type 
splits degenerate quark states in the spin and isospin states.  
Such a point of view provides an interesting way to consider structure of 
not only  three valence quark states but also multi-quark states.  

Let us start with a slightly general equation of motion of quarks under the 
influence of a potential generated by the chiral field (pion):  
\beq
\left(
i \dslash - g(\sigma (\vec x) + i \vec \tau \cdot \vec 
\pi(\vec x) \gamma_5 ) \right)
\psi = 0 \, .
\label{eqquark}
\eeq
where $\sigma (\vec x)$ and $\vec \pi (\vec x)$ are static 
scalar sigma 
and pseudoscalar pion fields which are regarded as
background potentials for the quarks. 
In order to clarify the role 
of the pion, let us assume the hedgehog configuration in which isospin of the 
pion field  $\vec \pi(\vec x)$ is correlated to the spatial orientation $\hat r$, 
$\vec \pi (\vec x)=\hat r h(r)$.
Here  $h(r)$ (and  $\sigma(\vec x) \to \sigma(r)$)  
are spherically symmetric functions. 
The equation of
motion with this pion (and sigma) background field affects the motion of the
light $u,d$ quarks, but not that of the strange $s$ quark.  
At this point, we break ``spontaneously" the flavor SU(3) symmetry.  

Write the four component spinor for the lowest $s$ state,
\beq
\psi =
\left(
\begin{array}{c}
u(r) \\
- i \vec \sigma \cdot \hat r v(r)
\end{array}
\right) \chi \, ,
\eeq
where $\chi$ is a spin-isospin spinor.  
Then after performing spatial integral the hamiltonian takes on
the  form:
\beq
H = c_1
+ c_2 \bra \chi | \vec \sigma \cdot \vec \tau | \chi \ket \, , 
\label{hamiltonian}
\eeq
where the first term is from the kinetic and $\sigma$ (mass) terms of 
(\ref{eqquark}), and 
the second term from the pion term, 
with $c_1$ and $c_2$ being numerical constants.  
A crucial point here is the appearance of the spin and isospin 
coupling $\sigma \tau$ term.
Without that term, 
there is a four fold degeneracy in the spin-isospin
states, $u\uparrow$, $u\downarrow$,$d\uparrow$ and $d\downarrow$.
Now if the $\sigma \tau$ term is turned on, then a conserved  quantity
is the grand angular momentum, the sum of the total angular momentum   
$\vec j = \vec l + \vec s$ and isospin $\vec t$, $\vec K = \vec j + \vec t$.  
When $\vec l = 0$, the above four spin and isospin states form
four eigenstates of $\vec K$ with 
$(K, K_z) = (0,0), (1,-1), (1,0), (1, 1)$.  
Since the coefficient $c_2$ is positive for the system of the baryon number 
$B = 1$, the $K=0$ state
is lowered whereas the $K=1$ state is pushed up.  
In the following discussion, we denote eigenstates of $K$ by $h$ instead of 
$u\uparrow$ and etc.  

\begin{figure}
       \centerline{\includegraphics[width=6cm]
                                   {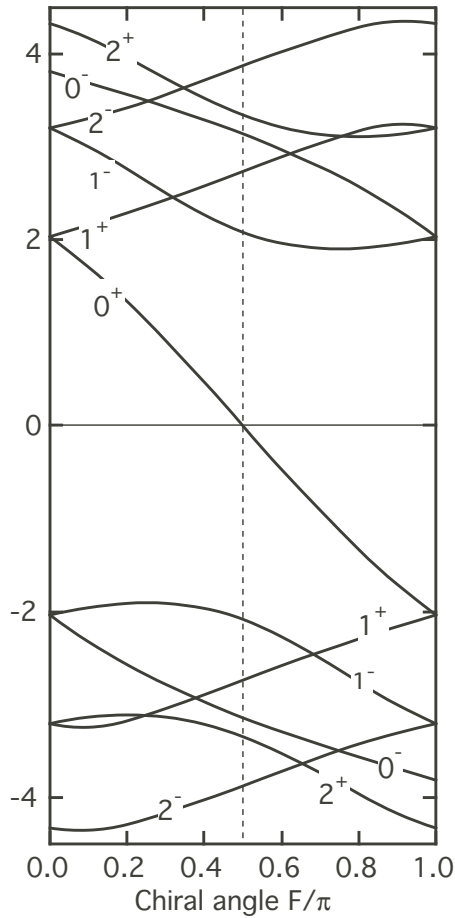}}
\centering
\begin{minipage}{12cm}
   \caption{\small 
   Eigenenergies of the hedgehog quark in the chiral bag model
   as functions of the chiral angle $F$. 
   }
   \label{eigenall}
 \end{minipage}
\end{figure}

In order to see this point slightly in a quantitative manner, we
adopt the result from the chiral bag model, where quarks are
confined in a spherical bag of a radius $R$, and  interact with the pion
($\sigma$ and $\pi$) at the bag surface. 
A mean field solution is then obtained in the hedgehog configuration.  
For instance, eigenergies of hedgehog quarks are 
obtained as functions of the pion strength (chiral angle) at
the bag surface $F(r=R)\equiv F$. 
Here $F$ is defined as a polar angle of the sigma and pion field, 
$\tan F = h(R)/\sigma(R)$~\cite{cbag}.  
In Fig.~\ref{eigenall}, energy levels for quarks are
shown as functions of $F$ for several lower lying states in units
of $1/R$.  
The labels denote $K^P$, where $P$ is parity.  
When $F = 0$, there is degeneracy in the states of $K^P$ and $(K+1)^P$.  
As $F$ is turned on, the degeneracy is resolved.  
Furthermore, there is a reflection symmetry between the quark and antiquark 
levels, 
$E_{\rm quark}(F, K^P) = E_{\rm antiquark}(\pi -F, K^{P})$
with respect to the axis  $F = \pi /2$.  
At this angle (the magic angle) there appears precisely a zero mode of $0^+$.  

Now, for finite $F$, two energy levels start to split;  for instance,  
as $F$ is increased, $1^+$ level goes up, while $0^+$ level goes down.  
It is a general feature for degenerate pairs at $F = 0$ that 
$E(F, K^P) <  E(F, (K+1)^P)$.  
Up to   $F \lsim 1$, the ordering of the lower three 
energy levels is
\beq
E_{0^+}  <  E_{1^+}  <  E_{1^-}  \, .
\eeq
However, at  $F \sim 1$  the level crossing occurs, and for $F \gsim 1$ 
\beq
E_{0^+}  <   E_{1^-}  <  E_{1^+} \, .
\eeq
This change in the level ordering is crucial 
when we consider the structure of various baryons including  
$\Theta^+$.


Let us first see implications of the level crossing for $\Theta^+$.  
A little bit more quantitative discussions will be made shortly.  
The antistrange quark $\bar s$ is not subject to the chiral potential and 
stays in the $J^P = 1/2^+$ orbit determined by the MIT bag boundary condition 
($F =0$)~\cite{MIT}, 
$E_{\bar s} \sim 2.3/R$ for $m_{\bar s} \sim 150$ MeV and 
$R \sim 0.7$ fm.  
Contrary, the light $h$ ($u,d$) quarks are influenced by $F$.  
For finite $F$, three $h$ quarks occupy the lowest 
$K^P = 0^+$ level due to the color degrees of freedom.  
Now the fourth $h$ quark enters the next level, which is 
$1^+$ for $F \lsim 1$ and $1^-$ for $F \gsim 1$.  
Depending on which orbit the fourth quark is in, the total parity of 
$\Theta^+$ changes; 
for $F \lsim 1$ $P = -$, while for $F \gsim 1$ $P = +$.  
The predictions of the chiral quark model and the Skyrme model 
correspond to the latter case.  

To make further discussions in a slightly quantitative manner, 
we consider a configuration which was found to 
be optimal for nucleon structure, where 
$F \sim \pi/2$ and $R \sim -0.7$ fm~\cite{cbag}.  
In this case the unit of  energy  is $1/R \sim 300$ MeV.  
In the following discussions, energy is estimated for the hedgehog 
configuration.  
Baryon states of the proper spin and isospin should be obtained 
by performing spin and isospin projection.  
The parity of the projected state is, however, not changed from that of the 
hedgehog.

\noindent
\underbar{Ground state nucleon $(S = 0)$}

Three hedgehog quarks ($h$) occupy the 
lowest $K^P = 0^+$ states.  
This quark configuration is denoted as $(0^+_h)^3$.  
At around $F \sim \pi/2$, the energy of the $0^+$ state is almost zero, and hence
the energy contribution from the valence quarks is also almost zero.  
However,  energies are also supplied from  the vacuum as the Casimir 
energy and in the soliton  field  outside the bag.  
The total energy of the hedgehog configuration is then about 1.3 GeV.  
In the chiral bag model, when using the experimental value for the 
pion decay constant  the  hedgehog mass is larger than the experimental 
masses.  
Also, for physical nucleons, spin and  isospin projection has to be 
performed, which adds a rotational energy as proportional to 
$I(I+1)$, where $I$ is the isospin value of the nucleons.  
Hence the mass of the ground state nucleon is slightly larger than 
the mass of the hedgehog  which is about 1.4 GeV.  

\noindent
\underbar{$\Theta^+ (S = 1)$:}

As we have shown, a tentative quark configuration of $\Theta^+$ is 
$(0^+_h)^3(1^-_h)^1(1/2^-_{\bar s})^1$.  
Here the parity of the antiquark $\bar s$ includes the intrinsic one.  
The excitation energy of the pentaquark state is about 
$4/R$, which is about 1.2 GeV.  
Therefore, in a rough estimation, the mass of the 5 quark state is about 
twice (or slightly less) of the mass of the ground state.  
If the nucleon mass is normalized to the experimental value of 
938 MeV, then the mass of the pentaquark state would be about 1.8 GeV.  
This value is larger as compared to the observed mass of 
$\Theta^+$.  
However, we do not yet include the expected hyperfine 
splitting~\cite{diakonov}.  
In fact, 1.8 GeV is close to the mean value of the mass of the antidecuplet 
in the chiral quark soliton model.  
From the quark energy level of Fig.~\ref{eigenall}, we  expect that 
an excited state of five quarks 
may exist  slightly above  $\Theta^+$ with 
having negative parity.  

\noindent
\underbar{Excited states of the nucleon $(S = 0)$}

By lifting one $h$ quark into a higher level, we can form 
excited states.  
At $F \sim \pi/2$, the first excited level is $1^-$ and 
the next one is $1^+$.  
Hence the relevant three quark configurations are 
$(0^+_h)^2 (1^-_h)^1$ and  $(0^+_h)^2 (1^+_h)^1$, respectively.  
If we take this result seriously, the first excited state of the nucleon 
would be $1/2^-$ with a mass about $2/R \sim 600$ MeV above 
the ground state, while the second  excited state is $1/2^+$ at about a few hundreds 
MeV above the $1/2^-$ state.  
If we make a rough argument in which a slightly larger bag and small 
$F$ would be preferred for excited states due to less number of (two) $h$ quarks 
in the lowest $0^+$, 
then the above conclusion would be slightly modified; 
the ordering of $1/2^{\pm}$ states would change.  
Qualitatively, we expect that the two $1/2^{\pm}$ states appear with 
similar excitation energies.  
We may identify them with  $N(1440)$ and  $N(1535)$.  
The nature of $N(1440)$ is, however, quite different from 
an ordinary picture of radial excitation.  
Here, the energy splitting is produced by the interaction with the 
chiral potential.  
It would be interesting to consider such a configuration for the positive parity 
$N(1440)$ resonance.  
On the other hand, the $1^-$ state is an $l = 1$ orbital  excitation 
just as in an ordinary interpretation.  

\noindent
\underbar{
$\Lambda(g.s.)$ and  $\Lambda({\rm excited \; \; states})$, $(S = -1)$}

\begin{figure}
       \centerline{\includegraphics[width=6cm]
                                   {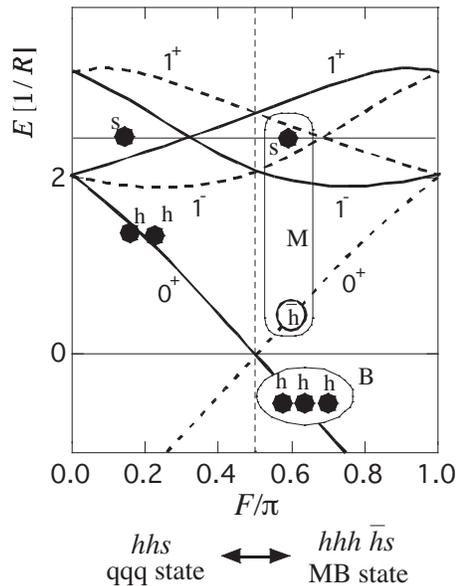}}
\centering
\begin{minipage}{12cm}
   \caption{\small Quark occupation for the $\Lambda$ particles.  
   Low lying quark 
   levels are shown for quarks (solid lines) and antiquarks (dashed lines).  }
   \label{mb_bound}
 \end{minipage}
\end{figure}

For the ground state, the three quark configuration is 
$(0^+)_h^2(1/2^+)_s^1$.  
At $F \sim \pi/2$, the energy appears to be higher than the ground state 
by $2/R \sim 600$ MeV.  
In a chiral bag calculation, however, larger hyperon ($hhs$) bag is preferred as compared to 
the nucleon ($hhh$) bag, which brings the mass of the ground state $\Lambda$ 
nearly at the right place~\cite{park}.  
For excited states, one $h$ jumps into the first excited orbit, 
either $0^+$ or $1^-$ state depending on the strength of the pion field.  
The excitation energy  is once again of order $2/R \sim $ 600 MeV 
(or smaller for larger $1/R$) above the ground state $\Lambda$.  
It is not possible to discuss more quantitative discussions 
such as for the splittings among the flavor singlet (predominantly 
$\lambda(1405)$), negative parity flavor octets 
($\Lambda(1670)$, $\Lambda(1690)$)
and  the one of positive parity ($\Lambda(1600)$).  
However, the following remarks may deserve being pointed out.  
When the pion field is strong ($F > \pi/2$), the quark 
configuration $(0^+)_h^2(1/2^+)_s^1$
for $\Lambda(g.s.)$
may be interpreted 
as a meson-baryon bound state (Fig.~\ref{mb_bound}).  
In this region, the two hedgehog quarks in the $0^+$ orbit 
dive into the negative energy see.  
This situation  may be interpreted as a vacuum of baryon number one 
(hedgehog soliton) and one anti-hedgehog ($\bar h$) in the $0^+$ 
orbit.  
The parity of the anti-hedgehog state here  includes both the intrinsic and 
orbital ones, and therefor, the orbital state of this $\bar h$ is $p$ state.  
The $p$ state $\bar h$ and the $s$ state $s$ quarks form an anti-kaon
($\bar K$) in a $p$ orbit.  
A similar interpretation is also possible for the excited $\Lambda$'s by 
lifting $\bar h$ into a higher orbit.  
A three quark state or meson-baryon 
state for $\Lambda$ is somewhat a matter of interpretation 
depending on the strength of the pion field.  

The conclusion made  in the above qualitative arguments, especially for the 
parity of the pentaquark state $\Theta^+$ is very important.  
This would be a clear  evidence in which the role of the pion is 
appreciated in explaining very fundamental  properties  
such as parity.  
In the experiments so far, parity of $\Theta^+$, as well as its spin, 
has not been determined yet.   
It is crucially important to know in the next step these quantum numbers 
in order to further understand the structure of the pentaquark state.  

Another important and interesting property of $\Theta^+$ 
is the narrow decay width $\Gamma \lsim 25$ MeV. 
This is once again in  agreement with the prediction of the 
chiral soliton model.  
The small width can be explained  by the small phase space 
in the $p$-wave coupling of $\Theta ^+\to K N$.  
The coupling constant extracted from the experimental width of 
25 MeV is about $g \sim 4$, in the same order of magnitude as 
other strong meson-baryon coupling constant, e.g. 
$g_{\pi NN} \sim 13$.  

The finding of $\Theta^+$ reminds us another recent finding of 
$D_s(2317)$ meson, whose quark content  is $c \bar s$~\cite{ds2317}.  
The properties of $\Theta^+$ and $D_s$ have some similarities in that; 
both of them contain anti-strangeness $\bar s$, their masses are significantly 
smaller than a naive quark model prediction, and their widths are unexpectedly 
small.  
For $\Theta^+$, we have argued that these properties may be 
explained by a strong interaction dynamics driven by the hedgehog pion, 
associated with the Nambu-Goldstone bosons of spontaneously broken chiral symmetry.  
Also for mesons, the importance of chiral symmetry in the presence of heavy quarks 
has been pointed out, giving a reasonable explanation for the properties 
of $D_s$ mesons~\cite{nowak,bardeen}.  
It will be interesting to pursue more the role of chiral symmetry  
for both meson and baryon dynamics.  


The pentaquark state is interesting in its own right, but also it opens
a step toward  multi-quark states.  
Further investigations on properties of the pentaquark baryons
should  shed  light on rich structure of hadronic matter.  

\vspace*{1cm}

\noindent
{\it Note after the submission}:  
After the submission of the manuscript, there appear 
several related papers.  
In Ref.~\cite{karliner} and ~\cite{jaffe}, a diquark-triquark and 
diquark-diquark-$\bar s$ configurations, respectively, were 
considered with color-magnetic interactions.  
Apart from the spin and parity assignment of $\Theta^+$, 
they predicted quite different pattern for other resonance states.  
In Ref.~\cite{zhu}, a QCD sum rule calculation predicted 
$J^P$ of $\Theta^+$ to be $1/2^-$.  
Further studies of exotic configurations will certainly be needed.  


\subsection*{Acknowledgments}
I would like to thank Takashi Nakano for illuminating discussions 
about the new finding  of the $\Theta^+$ particle.  
Thanks are also due to Hiroshi Toki for discussions on the 
role of the pion in hadron physics.

\end{document}